\documentclass[znote,zbstdefault]{./LaTeX/zeus/zeus_paper}
\usepackage[english]{babel}

\chardef\usc=95
\chardef\til=126
\catcode`\@=11 % @ signs are now treated as letters
\DeclareRobustCommand\xdotspace{\futurelet\@let@token\@xdotspace}
\def\@xdotspace{%
  \ifx\@let@token.\else
  \ifx\@let@token\bgroup.\else
  \ifx\@let@token\egroup.\else
  \ifx\@let@token\/.\else
  \ifx\@let@token\ .\else
  \ifx\@let@token~.\else
  \ifx\@let@token!.\else
  \ifx\@let@token,.\else
  \ifx\@let@token:.\else
  \ifx\@let@token;.\else
  \ifx\@let@token?.\else
  \ifx\@let@token/.\else
  \ifx\@let@token'.\else
  \ifx\@let@token).\else
  \ifx\@let@token-.\else
  \ifx\@let@token\@xobeysp.\else
  \ifx\@let@token\space.\else
  \ifx\@let@token\@sptoken.\else
   .\space
   \fi\fi\fi\fi\fi\fi\fi\fi\fi\fi\fi\fi\fi\fi\fi\fi\fi\fi}
\catcode`\@=12 % @ signs are no longer letters

\newcommand{\stru}[2]{%
   \relax\ifmmode\hbox{\vrule height#1 depth#2 width0pt}%
   \else\vrule height#1 depth#2 width0pt\fi}
\newcommand{\Ronum}[1]{\uppercase\expandafter{\romannumeral#1}}
\newcommand{\ronum}[1]{\expandafter{\romannumeral#1}}
\DeclareRobustCommand{\LaTeXZ}{%
  \LaTeX\kern-.05em4\kern-.1em
  {\raisebox{-0.2ex}{$\scriptstyle\text{ZEUS}$}}\xspace}
\DeclareMathAlphabet{\mathbf}{OT1}{cmr}{bx}{sl}
\newcommand{\eVdist}{\kern-0.06667em}

\newcommand{\mev}{{\,\text{Me}\eVdist\text{V\/}}}
\newcommand{\gev}{{\,\text{Ge}\eVdist\text{V\/}}}

\newcommand{\Tesla}{\,\text{T}}

\newcommand{\slashfrac}[2]{%
  \raisebox{0.5ex}{\ensuremath #1}\kern-0.12em/\kern-0.08em
  \raisebox{-.8ex}{\ensuremath #2}}
\newcommand{\sqr}[3]{%
    {\vcenter{\hrule height.#3ex\hbox{\vrule width.#2ex height#1ex
     \kern#1ex\vrule width.#3ex}\hrule height.#2ex}}}

\catcode`\@=11 % @ signs are now treated as letters
\newcommand{\parenbar}{\mathpalette\p@renb@r}
\def\p@renb@r#1#2{\vbox{%
  \ifx#1\scriptscriptstyle \dimen@.7em\dimen@ii.2em\else
  \ifx#1\scriptstyle \dimen@.8em\dimen@ii.25em\else
  \dimen@1em\dimen@ii.4em\fi\fi \offinterlineskip
  \ialign{\hfill##\hfill\cr
    \vbox{\hrule width\dimen@ii}\cr
    \noalign{\vskip-.3ex}%
    \hbox to\dimen@{$\mathchar300\hfil\mathchar301$}\cr
    \noalign{\vskip-.3ex}%
    $#1#2$\cr}}}
\catcode`\@=12 % @ signs are no longer letters

\newcommand{\IP}{{\rm I$\kern-0.01667em$P}\xspace}

\mathchardef\qsm=63
\mathchardef\pls=43
\mathchardef\mns=512
\mathchardef\plm=518
\mathchardef\eql=61
\mathchardef\smallleft=300
\mathchardef\smallright=301
\mathchardef\les=316
\mathchardef\gre=318
\mathchardef\leq=532
\mathchardef\grq=533
\catcode`\@=11 % @ signs are now treated as letters
\newcounter{pict@width}
\newcounter{pict@height}
\newlength{\pict@scale}
\newcommand{\psfigadd}[4]{%
\setcounter{pict@width}{1*\ratio{#2+\pict@scale/2}{\pict@scale}}
\setcounter{pict@height}{1*\ratio{#3+\pict@scale/2}{\pict@scale}}
\setlength{\unitlength}{\pict@scale}
\hbox to #2{\hspace{-\fill}\begin{picture}(\thepict@width,\thepict@height)
\put(0,0){\psfig{figure=#1,width=#2,height=#3,clip=}}
\SetScale{0.283466457}
\SetWidth{1.763889}
{#4}
\end{picture}}
}
\newcounter{pict@widthfst}
\newcounter{pict@widthscd}
\newcounter{pict@widthtot}
\newcommand{\psfigaddtwo}[7]{%
\setcounter{pict@widthfst}{1*\ratio{#2+\pict@scale/2}{\pict@scale}}
\setcounter{pict@widthscd}{1*\ratio{#2+#4+\pict@scale/2}{\pict@scale}}
\setcounter{pict@widthtot}{1*\ratio{#2+#4+#6+\pict@scale/2}{\pict@scale}}
\setcounter{pict@height}{1*\ratio{#3+\pict@scale/2}{\pict@scale}}
\setlength{\unitlength}{\pict@scale}
\hbox{\hspace{-\fill}\begin{picture}(\thepict@widthtot,\thepict@height)
\put(0,0){\psfig{figure=#1,width=#2,height=#3,clip=}}
\put(\thepict@widthscd,0){\psfig{figure=#5,width=#6,height=#3,clip=}}
\SetScale{0.283466457}
\SetWidth{1.763889}
{#7}
\end{picture}}
}
\newcommand{\psfigror}[4]{%
\setcounter{pict@width}{1*\ratio{#2+\pict@scale/2}{\pict@scale}}
\setcounter{pict@height}{1*\ratio{#3+\pict@scale/2}{\pict@scale}}
\setlength{\unitlength}{\pict@scale}
\hbox{\begin{picture}(\thepict@width,\thepict@height)
\put(0,\thepict@height){\psfig{figure=#1,width=#3,height=#2,clip=,angle=270}}
\SetScale{0.283466457}
\SetWidth{1.763889}
{#4}
\end{picture}}
}
\newcommand{\psfigrol}[4]{%
\setcounter{pict@width}{1*\ratio{#2+\pict@scale/2}{\pict@scale}}
\setcounter{pict@height}{1*\ratio{#3+\pict@scale/2}{\pict@scale}}
\setlength{\unitlength}{\pict@scale}
\hbox{\begin{picture}(\thepict@width,\thepict@height)
\put(0,0){\psfig{figure=#1,width=#3,height=#2,clip=,angle=90}}
\SetScale{0.283466457}
\SetWidth{1.763889}
{#4}
\end{picture}}
}
\catcode`\@=12 % @ signs are no longer letters
\newlength\listtextwidth

\catcode`\@=11 % @ signs are now treated as letters
\newlength{\@tabfninsert}
\newlength{\@tabfnwidth}
\newcommand{\tabfootnote}[2]{%
  \setlength{\@tabfninsert}{0.8em}
  \setlength{\@tabfnwidth}{\textwidth}
  \addtolength{\@tabfnwidth}{-\@tabfninsert}
  \addtolength{\@tabfnwidth}{-0.4em}
  \noindent\makebox[\@tabfninsert][r]{\footnotesize$^{#1}$\hfil}\hfill%
  \parbox[t]{\@tabfnwidth}{\footnotesize #2\hfill}}
\catcode`\@=12 % @ signs are no longer letters

\def\Cref#1{Chapter~\ref{#1}}                   % ref to chapters
\def\cref#1{chapter~\ref{#1}}
\def\citeCTD{{\cite{%
nim:a279:290,*npps:b32:181,*nim:a338:254%
}}\xspace}
\def\citeMVD{{\cite{%
nim:a581:656%
}}\xspace}
\begin{document}

\prepnum{DESY--09--059}

\title{
Scaled momentum distributions of charged particles in dijet
photoproduction at HERA
}                                                       
                    
\author{ZEUS Collaboration\\}

\date{21st April, 2009}

\abstract{
The scaled momentum distributions of charged particles in jets have been measured for dijet photoproduction with the ZEUS detector at HERA using an integrated luminosity of 359 pb$^{-1}$.  The distributions are compared to predictions based on perturbative QCD carried out in the framework of the modified leading-logarithmic approximation (MLLA) and assuming local parton-hadron duality (LPHD). The universal MLLA scale, $\Lambda_{\rm eff}$, and the LPHD parameter, $\kappa^{\rm ch}$, are extracted.
}
\makezeustitle

\begin{center}                                                                                     
{                      \Large  The ZEUS Collaboration              }                               
\end{center}                                                                                       
  S.~Chekanov,                                                                                     
  M.~Derrick,                                                                                      
  S.~Magill,                                                                                       
  B.~Musgrave,                                                                                     
  D.~Nicholass$^{   1}$,                                                                           
  \mbox{J.~Repond},                                                                                
  R.~Yoshida\\                                                                                     
 {\it Argonne National Laboratory, Argonne, Illinois 60439-4815, USA}~$^{n}$                       
\par \filbreak                                                                                     
  M.C.K.~Mattingly \\                                                                              
 {\it Andrews University, Berrien Springs, Michigan 49104-0380, USA}                               
\par \filbreak                                                                                     
  P.~Antonioli,                                                                                    
  G.~Bari,                                                                                         
  L.~Bellagamba,                                                                                   
  D.~Boscherini,                                                                                   
  A.~Bruni,                                                                                        
  G.~Bruni,                                                                                        
  F.~Cindolo,                                                                                      
  M.~Corradi,                                                                                      
\mbox{G.~Iacobucci},                                                                               
  A.~Margotti,                                                                                     
  R.~Nania,                                                                                        
  A.~Polini\\                                                                                      
  {\it INFN Bologna, Bologna, Italy}~$^{e}$
% ...
\newpage                                                                                           
\enlargethispage{5cm}                                                                              
$^{\    1}$ also affiliated with University College London,                                        
United Kingdom\\                                                                                   
% ...
$^{\    5}$ now at Institute of Aviation, Warsaw, Poland \\                                        
% ...
%                                                                                                  
% \par         % if index listing & table fit to 1 page, put gap here                              
\newpage   % alternatively: go to newpage, if page is too small                                    
                                                           %                                       
% \institute_references_start    % do not touch or move this line !                                
                                                           %                                       
\begin{tabular}[h]{rp{14cm}}                                                                       
$^{a}$ &  supported by the Natural Sciences and Engineering Research Council of Canada (NSERC) \\  
$^{b}$ &  supported by the German Federal Ministry for Education and Research (BMBF), under        
          contract Nos. 05 HZ6PDA, 05 HZ6GUA, 05 HZ6VFA and 05 HZ4KHA\\                            
$^{c}$ &  supported in part by the MINERVA Gesellschaft f\"ur Forschung GmbH, the Israel Science   
          Foundation (grant No. 293/02-11.2) and the US-Israel Binational Science Foundation \\    
$^{d}$ &  supported by the Israel Science Foundation\\                                             
$^{e}$ &  supported by the Italian National Institute for Nuclear Physics (INFN) \\                
$^{f}$ &  supported by the Japanese Ministry of Education, Culture, Sports, Science and Technology 
          (MEXT) and its grants for Scientific Research\\                                          
$^{g}$ &  supported by the Korean Ministry of Education and Korea Science and Engineering          
          Foundation\\                                                                             
$^{h}$ &  supported by the Netherlands Foundation for Research on Matter (FOM)\\                   
$^{i}$ &  supported by the Polish State Committee for Scientific Research, project No.             
          DESY/256/2006 - 154/DES/2006/03\\                                                        
$^{j}$ &  partially supported by the German Federal Ministry for Education and Research (BMBF)\\   
$^{k}$ &  supported by RF Presidential grant N 1456.2008.2 for the leading                         
          scientific schools and by the Russian Ministry of Education and Science through its      
          grant for Scientific Research on High Energy Physics\\                                   
$^{l}$ &  supported by the Spanish Ministry of Education and Science through funds provided by     
          CICYT\\                                                                                  
$^{m}$ &  supported by the Science and Technology Facilities Council, UK\\                         
$^{n}$ &  supported by the US Department of Energy\\                                               
$^{o}$ &  supported by the US National Science Foundation. Any opinion,                            
findings and conclusions or recommendations expressed in this material                             
are those of the authors and do not necessarily reflect the views of the                           
National Science Foundation.\\                                                                     
$^{p}$ &  supported by the Polish Ministry of Science and Higher Education                         
as a scientific project (2006-2008)\\                                                              
$^{q}$ &  supported by FNRS and its associated funds (IISN and FRIA) and by an Inter-University    
          Attraction Poles Programme subsidised by the Belgian Federal Science Policy Office\\     
$^{r}$ &  supported by an FRGS grant from the Malaysian government\\                               
\end{tabular}                                                                                      
                                                           %                                       
% \institute_references_end     % do not touch or move this line !                                 
                                                           %                                       
%\end{document}                                                                                     

\section{Introduction}

The formation of jets of hadrons can be described as a convolution of parton showering and hadronisation. Within perturbative QCD (pQCD), the parton shower can be described as long as the energy scale involved is sufficiently above the intrinsic scale of QCD, $\Lambda_{\rm QCD}$. Hadronisation describes the process by which coloured partons become confined in colour-neutral hadrons.  It cannot be described within pQCD.

Perturbative QCD calculations can be performed using matrix elements up to a certain order in the strong coupling constant, $\alpha_s$.  Alternatively, a resummation approach can be adopted, such as the modified leading-logarithmic approximation (MLLA)~\cite{MLLA1,*MLLA2,*MLLA3,*MLLA4,*Bk:dokshitzerQCD,*MLLA5}, where in addition to the fixed-order matrix elements, a subset of dominant terms of all orders in $\alpha_s$ are included. In particular, pQCD based on the MLLA can be used to predict the multiplicity and momentum spectra of partons produced within cones centred on the initial parton direction. The MLLA may only be used to describe partons at scales above some minimum cutoff, $\Lambda_{\rm eff}>\Lambda_{\rm QCD}$. The value of $\Lambda_{\rm eff}$ is predicted to be independent of the process considered. The local parton hadron duality (LPHD)~\cite{LPHD1} hypothesis predicts that charged-hadron distributions should be related to the predicted parton distributions by a constant normalisation scaling factor, $\kappa^{\rm ch}$. 

Tests of the MLLA have been performed before using data from $e^{+}e^{-}$ collisions at LEP~\cite{Akrawy:1990ha,Adeva:1991it} and PETRA~\cite{Braunschweig:1990yd}, deep inelastic scattering (DIS) $ep$ collisions at HERA~\cite{Derrick:1995ca,Breitweg:1999nt}, (anti-) neutrino-nucleon interactions from the NOMAD experiment~\cite{Altegoer:1998py} and $p\bar{p}$ collisions at the Tevatron~\cite{Acosta:2002gg}.  In this analysis, the multiplicity and momentum spectra of charged hadrons within jets are studied using photoproduction ($\gamma p$) in $ep$ collisions, in which a quasi-real photon emitted from the incoming electron collides with a proton. The events were required to have two and only two reconstructed jets and the sample was enriched in events in which the photon interacted electromagnetically as a point-like particle. The analysis probes energy scales in the range $19$ to $38$\gev, which spans the energy region between those accessed previously by the ZEUS, using $ep$ DIS collisions~\cite{Derrick:1995ca,Breitweg:1999nt}, and CDF collaborations~\cite{Acosta:2002gg}. The quantities $\Lambda_{\rm eff}$ and $\kappa^{\rm ch}$ are extracted and their universality tested.

%The analysis~\cite{thesis:john} described here is the first of its kind using photoproduction in $ep$ collisions, in which a quasi-real photon emitted from the incoming electron collides with a proton.  Since the expected lifetime of the quasi-real exchanged photon is long when compared to the duration of its interaction with the proton, it is valid to view the reaction as a $\gamma p$ collision. 
%The relative expected longevity of the quasi-real exchanged photon, compared to the typical time-scale of the interaction, implies that it is valid to view the reaction as a $\gamma p$ collision. 

%%%%%%%%%%%%%%%%%%%%%%%%%%%%%%%%%%%%%%%%%%%%%%%%%%%%%%%%%%%%%%%%%%%%%%%%%%%%%%

\section{The MLLA framework}
\label{sec:MLLAmethod}

The MLLA includes all terms of order $\alpha_s^n\log^{2n}(E_{\rm init}^{\rm pl})$ and $\alpha_s^n\log^{2n-1}(E_{\rm init}^{\rm pl})$, where $n$ is the set of positive integers and $E_{\rm init}^{\rm pl}$ is the energy of the initial outgoing parton in the centre-of-mass frame of the incoming struck parton and exchanged photon.  The ``${\rm pl}$'' superscript denotes a parton-level quantity.  The MLLA accounts for colour-coherence effects between diagrams of the same order of $\alpha_s$ by enforcing an angular-ordering scheme~\cite{angular}. 

The MLLA equations describe the momentum and multiplicity spectra of partons at a specified scale, $Q_0$.  They are only strictly valid for partons satisfying $x_p^{\rm pl}=|p_p^{\rm pl}|/E_{\rm init}^{\rm pl}\ll1$, where $p_p^{\rm pl}$ is the 3-momentum of a parton in the centre-of-mass frame. For the MLLA predictions used here, the singularities were regularised by a single $p_T^{\rm rel,pl}$ cut-off at scale $Q_0>\Lambda_{\rm eff}$, where $p_T^{\rm rel,pl}$ is the transverse momentum with which the parton was emitted with respect to its parent. This is not the only possible way to regularise the MLLA; other forms lead to different predictions, particularly at low $x_p^{\rm pl}$~\cite{Lupia:1997hj}.  

Predictions at the lowest valid scale, $Q_0=\Lambda_{\rm eff}$, give the so-called limiting momentum spectrum of partons, $\bar{D}^{\rm lim,pl} = \frac{dN^{\rm pl}}{d\xi^{\rm pl}}$, where $\xi^{\rm pl}=\ln(1/x_p^{\rm pl})$ and $N^{\rm pl}$ is the multiplicity of partons produced within a cone of opening angle, $\theta_c^{\rm pl}$, measured with respect to the axis of the initial parton.  The predictions assume that $\theta_c^{\rm pl}$ is small.  The shape of the predicted spectrum is roughly Gaussian, although it falls rapidly to zero as $\xi^{\rm pl}\rightarrow \ln\left(E_{\rm init}^{\rm pl}\sin(\theta_c^{\rm pl})/\Lambda_{\rm eff}\right)$, a consequence of the regularisation scheme adopted.

For a gluon-initiated parton-level jet, the function is written as~\cite{Khoze:2000iq}
\begin{eqnarray*}
  \bar{D}^{\rm lim,pl}_{\rm g-jet} & = & F_{\rm nMLLA}\frac{4n_c}{b}\Gamma(B)\int_{-\pi/2}^{\pi/2}e^{-B\alpha}\left[\frac{\cosh\alpha+(1-2\zeta)\sinh\alpha}{\frac{4n_c}{b}Y\frac{\alpha}{\sinh\alpha}}\right]^{B/2} \cdot~~~~~~~~~~~~~~~~~~~~~~~~~~~~~~~~
\end{eqnarray*}
\vspace{-0.6cm}
\begin{eqnarray}
 ~~~~~~~~~~~~~~~~~~~~~~~~~~~~~~~~~~~~~~
I_B\left(\sqrt{\frac{16n_c}{b}Y\frac{\alpha}{\sinh\alpha}[\cosh\alpha+(1-2\zeta)\sinh\alpha]}\right)\frac{d\tau}{\pi},
  \label{equ:1}
\end{eqnarray}
where the parameters $b=9$ and $B=101/81$ are QCD constants defined in terms of the number of colours, $n_c=3$, and number of flavours, $n_f=3$. The symbols $\Gamma$ and $I_B$ denote the Gamma and $B^{th}$-order modified Bessel functions, respectively.  The other variables are defined as
\begin{equation}
Y=\ln(E_{\rm init}^{\rm pl}\sin(\theta_c^{\rm pl})/\Lambda_{\rm eff}) ,~~~ \zeta=1-\xi^{\rm pl}/Y ,~~~ \alpha=\tanh^{-1}(2\zeta-1)+i\tau,
\label{equ:2}
\end{equation}
and, in the MLLA, $F_{\rm nMLLA}= 1$. The value of $n_f=3$ was chosen. The MLLA assumes massless partons and does not include any mass threshold effects for heavy flavours.  When a value of $n_f$ larger than $3$ is used instead, the theory is observed to give a poorer description of this and other data sets~\cite{Khoze:2000iq}.

At leading order (LO), the peak position of the limiting momentum spectrum, $\xi^{\rm pl}_{\rm peak}$, is predicted to be at
%
%\begin{equation}
%\xi^{\rm pl}_{\rm peak}=\frac{1}{2}\left[\left(\sqrt{\ln\left(\frac{E^{\rm pl}_{\rm init}\sin(\theta_c^{\rm pl})}{\Lambda_{\rm eff}}\right)}+0.54\right)^2-0.87\right].
%\label{equ:2a}
%\end{equation}
%
\begin{equation}
\xi^{\rm pl}_{\rm peak}=\frac{1}{2}Y + \sqrt{cY} - c,
%
%\xi^{\rm pl}_{\rm peak} = \frac{1}{2}\left[\ln\left(E_{\rm jet}\sin\left(\theta_{c}\right)/\Lambda_{\rm eff}\right)\right] + \sqrt{c\left[\ln\left(E_{\rm jet}\sin\left(\theta_{c}\right)/\Lambda_{\rm eff}\right)\right]} - c,
%
%\xi^{\rm pl}_{\rm peak}=\frac{1}{2}\left[\ln\left(E_{\rm jet}\sin\left(\theta_{c}\right)\right)-\ln\left(\Lambda_{\rm eff}\right)\right] + \sqrt{c\left[\ln\left(E_{\rm jet}\sin\left(\theta_{c}\right)\right)-\ln\left(\Lambda_{\rm eff}\right)\right]} - c,
%
\label{equ:2a}
\end{equation}
where $c=0.29$.

The limiting spectrum of quark jets, $\bar{D}^{\rm lim,pl}_{\rm q-jet}$, is related to that of gluon jets according to
\begin{equation}
\bar{D}^{\rm lim,pl}_{\rm q-jet} = \frac{1}{r}\bar{D}^{\rm lim,pl}_{\rm g-jet},
\label{equ:3}
\end{equation}
where $r=N_{\rm g-jet}^{\rm pl}/N_{\rm q-jet}^{\rm pl}$ is the ratio of parton multiplicities in gluon- and quark-initiated jets.  In the MLLA, $r=C_A/C_F=9/4$ where $C_A$ and $C_F$ are the gluon and quark colour factors, respectively.

Photoproduction samples contain both gluon- and quark-initiated jets, in the fractions denoted by $\epsilon_{\rm g}$ and $\epsilon_{\rm q} = 1-\epsilon_{\rm g}$, respectively.  Thus, the limiting spectrum for partons in all jets can be parameterised as
\begin{equation}
\bar{D}^{\rm lim,pl} = \left(\epsilon_{\rm g}+\frac{1-\epsilon_{\rm g}}{r}\right)\bar{D}^{\rm lim,pl}_{\rm g-jet}.
\label{equ:4}
\end{equation}

Solutions to the MLLA evolution equations have also been made at so-called next-to-MLLA order. Each of these solutions partially accounts for orders not included in the equations above.  With the next-to-MLLA corrections, $F_{\rm nMLLA}$ and $r$ differ from their MLLA values and both have a weak dependence on $E_{\rm init}^{\rm pl}$.  Next-to-MLLA $F_{\rm nMLLA}$ and $r$ values have been used in this analysis in the same way as they were by the CDF collaboration~\cite{Acosta:2002gg}, wherein more details can be found.  Their values were taken from three different next-to-MLLA calculations~\cite{Lupia:1997in,*Dremin:1999ji,*Catani:1992tm}, which differ in the way the additional orders are accounted for, leading to some spread in the predicted $F_{\rm nMLLA}$ and $r$ values. Here, constant values of $F_{\rm nMLLA}=1.3\pm0.2$ and $r=1.6\pm0.2$ were used, with the theoretical uncertainties covering the spreads. 

The LPHD approximation relates the limiting momentum spectrum of partons to that of charged hadrons within jets, $\bar{D}^{\rm lim,ch}$, via
\begin{equation}
\bar{D}^{\rm lim,ch} = \kappa^{\rm ch}\bar{D}^{\rm lim,pl} = \kappa^{\rm ch}\left(\epsilon_{\rm g}+\frac{1-\epsilon_{\rm g}}{r}\right)\bar{D}^{\rm lim,pl}_{\rm g-jet} = K\bar{D}^{\rm lim,pl}_{\rm g-jet},
\label{equ:5}
\end{equation}
i.e. $K=\kappa^{\rm ch}\left(\epsilon_{\rm g}+(1-\epsilon_{\rm g})/r\right)$. Due to isospin invariance, $\kappa^{ch}$ is expected to be approximately $2/3$.

%%%%%%%%%%%%%%%%%%%%%%%%%%%%%%%%%%%%%%%%%%%%%%%%%%%%%%%%%%%%%%%%%%%%%%%%%%%%%%

\section{The analysis strategy}
\label{sec:analstrat}

To compare the parton-level MLLA predictions to measured hadron-level data, while assuming LPHD, each variable within the MLLA had to be estimated using a related hadron-level quantity.  The hadron-level estimator for $E_{\rm init}^{\rm pl}$ was chosen to be $E_{\rm jet}=M_{2j}/2$, where $E_{\rm jet}$ is the energy of either hadron-level jet in the dijet centre-of-mass frame and $M_{2j}$ is the invariant dijet mass. The quantity $p_p^{\rm pl}$ was estimated using the momenta of the charged hadrons, $p_{\rm trk}$.  The loss of the neutral hadrons is accounted for via the LPHD factor $\kappa^{\rm ch}$.  The MLLA variable $\theta_c^{\rm pl}$ was estimated using the opening angle of a cone measured with respect to the reconstructed jet axis, $\theta_c$.  Accordingly, the quantity $\bar{D}^{\rm lim,ch}$, given in Eq.~\ref{equ:5}, was estimated using the hadron-level multiplicity distribution of charged hadrons per jet, $N^{\rm ch}_{\rm jet}$, measured in bins of $E_{\rm jet}$ and in cones of varying $\theta_c$, differentially in $\xi=\ln\left(E_{\rm jet}/|p_{\rm trk}|\right)$. These ${\rm d}N^{\rm ch}_{\rm jet}/{\rm d}\xi$ distributions will be referred to as the $\xi$ distributions. 

%%%%%%%%%%%%%%%%%%%%%%%%%%%%%%%%%%%%%%%%%%%%%%%%%%%%%%%%%%%%%%%%%%%%%%%%%%%%%%

\section{Experimental setup}

The data analysed here were collected using the ZEUS detector during the 2005 to 2007 running periods, in which electrons\footnote{The word ``electron'' is used as a generic term for electrons and positrons.} were collided with protons with energies of $E_e=27.5$\gev~and $E_p=920$\gev, respectively, corresponding to a centre-of-mass energy, $\sqrt{s}=318$\gev.  The total sample corresponds to an integrated luminosity of $359\pm9{\rm~pb^{\rm -1}}$. A detailed description of the ZEUS detector can be found elsewhere~\cite{pl:b297:404,zeus:1993:bluebook}. A brief outline of the components most relevant to this analysis is given below.  

Charged particles were tracked in the central tracking detector (CTD)~\citeCTD, the microvertex detector (MVD)~\citeMVD and the straw-tube tracker (STT)~\cite{Fourletov:2004iu}. The CTD and MVD were operated in a magnetic field of $1.43\Tesla$ provided by a thin superconducting solenoid. The CTD drift chamber covered the polar-angle\footnote{The ZEUS coordinate system is a right-handed Cartesian system, with the $Z$ axis pointing in the proton beam direction, referred to as the ``forward direction'', and the $X$ axis pointing towards the centre of HERA. The coordinate origin is at the nominal interaction point.\xspace} region \mbox{$15^\circ<\theta<164^\circ$}. The MVD silicon tracker consisted of a barrel (BMVD) and a forward (FMVD) section. The BMVD provided polar-angle coverage for tracks with three measurements from $30^\circ$ to $150^\circ$. The FMVD extended the polar-angle coverage in the forward region to $7^\circ$.  The STT covered the polar-angle region \mbox{$5^\circ<\theta<25^\circ$}.

The high-resolution uranium--scintillator calorimeter (CAL)~\cite{nim:a309:77,*nim:a309:101,*nim:a336:23,*nim:a321:356} consisted of three parts: the forward, the barrel and the rear calorimeters. Each part was subdivided transversely into towers and longitudinally into one electromagnetic and either one (in the rear) or two (in the barrel and forward) hadronic sections. The smallest subdivision of the calorimeter was called a cell. The CAL relative energy resolutions, as measured under test-beam conditions, were $0.18/\sqrt{E}$ for electrons and $0.35/\sqrt{E}$ for hadrons, with $E$ in\gev. 

%%%%%%%%%%%%%%%%%%%%%%%%%%%%%%%%%%%%%%%%%%%%%%%%%%%%%%%%%%%%%%%%%%%%%%%%%%%%%%

\section{Event reconstruction}
\label{eventrecon}

A three-level trigger system was used to select events online~\cite{zeus:1993:bluebook,epj:c1:109,Gttsmith}. At the first two levels, general characteristics of photoproduction collisions were required and background from beam-gas events was rejected.  At the third level, jets were reconstructed by applying the $k_T$ cluster algorithm~\cite{Art:ktclus} to the CAL cells and a loose dijet selection was applied.  

In the offline analysis, the hadronic final state was reconstructed using energy-flow objects~\cite{Art:efo1, thesis:hefa} (EFOs), which were formed from a combination of track and calorimeter information.  This approach optimised the energy resolution and improved the one-to-one correspondence between the detector-level objects and the hadrons.  The EFOs were corrected to account for energy losses in the dead material and were forced to be massless by setting the energy component equal to the magnitude of the three-momentum. 

Jets were reconstructed from EFOs using the $k_T$ cluster algorithm~\cite{Art:ktclus} in the longitudinally invariant inclusive mode~\cite{Art:ktclusinc} using the $p_T$ recombination scheme and with the $R$ parameter set to $R=1$.  

Photoproduction events are characterised by the low virtuality, $Q^2$, of the exchanged photon. At LO, photoproduction can be categorised as being either direct, if the photon interacts as a point-like particle, or resolved, if it fluctuates into a partonic system, which then interacts with the proton. The LO direct photoproduction processes are boson gluon fusion, $\gamma g\rightarrow q \bar{q}$, and QCD Compton scattering, $\gamma q\rightarrow qg$.  Important kinematic variables are the inelasticity, $y$, and the fraction of the photon momentum transferred to the hadronic final state, $x_\gamma$.  The variable $x_\gamma$ can be approximated using the observable $x_\gamma^{\rm obs}$, defined for a dijet event as
\begin{equation}
x_\gamma^{\rm obs}=\frac{\sum_{i=1}^{\rm 2}E_{T}^{{\rm jet(}i{\rm)}}\exp(-\eta^{{\rm jet(}i{\rm)}})}{2yE_e},
\label{equ:xgammaobs}
\end{equation}

where $E_{T}^{\rm jet}$ and $\eta^{\rm jet}$ denote the jet transverse energy and pseudorapidity in the laboratory frame, respectively. A value of $x_\gamma^{\rm obs}$ approaching one indicates an event from a direct-like photoproduction process.

%%%%%%%%%%%%%%%%%%%%%%%%%%%%%%%%%%%%%%%%%%%%%%%%%%%%%%%%%%%%%%%%

\section{Event selection}
\label{eventselect}

To remove non-photoproduction events it was required that:
\begin{itemize}
\item the longitudinal position of the reconstructed vertex was in the range $|Z_{\rm vtx}|\le 40$~cm;
\item $0.2\le y_{\rm JB}\le 0.85$, where $y_{\rm JB}$ is the Jacquet--Blondel estimator~\cite{proc:epfacility:1979:391} of $y$;
\item no scattered electron was observed in the CAL with $E_e^\prime>5$\gev and $y_e<0.85$, where $E_e^\prime$ is the energy of the scattered electron and $y_e$ is the electron-method estimator of $y$~\cite{emethod};
\item $P_T^{\rm miss}/\sqrt{E_T}\le 2\gev^{1/2}$, where $P_T^{\rm miss}$ and $E_T$ are the reconstructed missing and total transverse momenta, respectively;
\item $|t^{\rm top}_{\rm CAL}-t^{\rm bot}_{\rm CAL}|<6$~ns, where $|t^{\rm top}_{\rm CAL}-t^{\rm bot}_{\rm CAL}|$ is the difference between the arrival times of the first signals in the top and bottom halves of the CAL;
\item $N_{\rm trk}^{\rm pri}/N_{\rm trk}>0.1$, where $N_{\rm trk}^{\rm pri}/N_{\rm trk}$ is the ratio of the number of tracks fitted to the primary vertex to the total number of all tracks.
\end{itemize}
To select an exclusive dijet sample enriched in direct events it was required that:
\begin{itemize}
\item two jets were found such that:
\begin{itemize}
\item the highest $E_T^{\rm jet}$ jet, labelled 1, had $|\eta^{\rm jet1}|\le1$ and $E_T^{\rm jet1}\ge 17~\rm{\gev}$;
\item the second jet, labelled 2, had $|\eta^{\rm jet2}|\le1$ and $E_T^{\rm jet2}/E_T^{\rm jet1}\ge 0.8$;
\item the first and second jets satisfied $|\phi^{\rm jet1}-\phi^{\rm jet2}|\ge 0.9\pi$, where $\phi^{\rm jet}$ denotes the azimuthal angle of the jet;
\end{itemize}
\item no third jet was found with $|\eta^{\rm jet3}|\le2.4$ and $E_T^{\rm jet3}\ge 6~\rm{\gev}$;
\item $x_\gamma^{\rm obs}\ge 0.75$.
\end{itemize}
To ensure that the tracks were well reconstructed and not associated with secondary charged particles generated via nuclear interactions within the detector material it was required that:
\begin{itemize}
\item the track transverse momentum was greater than 150\mev;
\item the track pseudorapidity was between $\pm1.7$;
\item the track passed through at least 3 CTD super layers;
\item the track was associated to the primary vertex.
\end{itemize}

The requirement that there be two and only two jets roughly balancing in $E_T^{\rm jet}$ and in opposite hemispheres ensured that the events were LO-like, where the energy scale is well estimated using $M_{2j}/2$.  The $x_\gamma^{\rm obs}$ criterion was applied to minimise the influence of multi-parton interactions (MPIs)~\cite{Art:forshawMPI,Art:forshawMPI2, Art:srivastavaMPI}, which generate additional final-state hadrons and can disrupt the correspondence between the MLLA predictions and the data. After all the above selection, the data sample contained 23,449 events.

%%%%%%%%%%%%%%%%%%%%%%%%%%%%%%%%%%%%%%%%%%%%%%%%%%%%%%%%%%%%%%%%%%%%%%%%%%%%%%

\section{Acceptance corrections}
\label{sec:unfold}

Effects due to the limited detector and trigger acceptance, efficiency and resolution were corrected for in the data using a sample of events generated with the {\sc Pythia} MC model~\cite{Art:pythia, *Art:pythia2}.  The direct and resolved photoproduction processes were generated separately and combined in the ratio that best fit the $x_\gamma^{\rm obs}$ distribution in the data.  The {\sc Pythia} model includes the LO ($2\rightarrow2$) matrix elements, approximates higher-order processes using initial-state and final-state parton showers and simulates hadronisation using the Lund string model~\cite{Art:pytHad1}.  The CTEQ5L~\cite{Art:cteq5l} and GRV-G LO~\cite{Art:grvgloTN} parameterisations were used to describe the proton and photon PDFs, respectively.  The main sample included MPIs, simulated using the  ``simple model''~\cite{Art:pythia, *Art:pythia2} within {\sc Pythia}, although the effects from MPIs were predicted to be negligible in the final sample.  The detector simulation was based on {\sc Geant} 3.21~\cite{unp:geant} and included a complete simulation of the three-level trigger system.

The data were corrected bin-by-bin to the hadron-level using factors extracted from the MC equal to the ratio of the predicted hadron- to detector-level cross sections.  Here, the hadron-level was defined to contain all particles with an average lifetime greater than 0.01~ns.  % but not their decay products. 
The size of the bin-by-bin corrections were typically around 1.5.

The normalisation of the $\xi$ distributions was set such that the integral of the distributions over the full $\xi$ range equalled $\left<N^{\rm ch}_{\rm jet}\right>$, where $\left<N^{\rm ch}_{\rm jet}\right>$ denotes the average hadron-level charged-particle multiplicity within jets, with the appropriate cone and energy scale criteria applied.  The values of $\left<N^{\rm ch}_{\rm jet}\right>$ were extracted from the data by measuring the corresponding charged multiplicity distributions.  These were corrected to the hadron-level using unfolding matrices derived from the {\sc Pythia} MC sample.  Full details of the procedure are described elsewhere~\cite{Derrick:1995ca,thesis:tim}.

%%%%%%%%%%%%%%%%%%%%%%%%%%%%%%%%%%%%%%%%%%%%%%%%%%%%%%%%%%%%%%%%%%%%%%%%%%%%%%

\section{Systematic uncertainties}
\label{sec:syst}
A detailed study~\cite{thesis:john} of the sources of systematic uncertainty associated with the measurement was performed.  The dominant sources contributing to the systematic uncertainty on the $\xi$ distributions are listed below (the numbers in parentheses refer to the maximum uncertainty observed in any one bin):
\begin{itemize}
\item the $\pm3\%$ uncertainty in the CAL energy scale, propagated to the $\xi$ distributions by varying the CAL energies in the MC simulation accordingly ($\pm4\%$);
\item the uncertainty simulating nuclear interactions in the detector material and the production of charged secondary particles.  This was propagated to the $\xi$ distributions by varying the difference between the number of tracks gained and lost due to such effects in the MC by a factor of 2 ($\pm4\%$);
\item the uncertainty in the tracking efficiency, propagated to the $\xi$ distributions using the procedure described below ($+5\%$).
\end{itemize}
 
The MC slightly overestimated the number of tracks in the data, probably due to either the uncertainty in the hadronisation model or to inadequacies in the detector simulation. The unfolding procedure is only strongly sensitive to the detector-level simulation rather than the hadron-level MC model and it was assumed that this was the sole cause of the excess.  This systematic uncertainty was evaluated by randomly failing detector-level tracks in the MC with track rejection rates evaluated in bins of $E_{\rm jet}$, $\theta_{\rm trk}$ and $1/p_{\rm trk}$, where $\theta_{\rm trk}$ is the polar angle between the track and the jet axis. The largest rejection rate was 14\%. The analysis was then repeated and the resulting difference in the $\xi$ distributions was included in the systematic uncertainty. All the systematic uncertainties were added in quadrature.

In the next section, several fits of the data are discussed. While nominally fitting the data and when evaluating the associated $\chi^2$ values, only the statistical uncertainties were considered.  The systematic uncertainties on the data were propagated, however, to the fitted parameters using the ``offset method''.  To apply the ``offset method'', the fit is repeated for each source of systematic uncertainty, shifting the nominal data by the uncertainty attributed to that one source.  The differences between the values of the parameters extracted from the nominal and the shifted data are then summed in quadrature and included as the total systematic uncertainty on the parameter itself. 

%The measured data were fit a number of times, as discussed in the next section.  Only the statistical uncertainty was considered while nominally fitting the data points and when evaluating the $\chi^2$ values associated with the fits.  The systematic uncertainties on the data were propagated, however, to the fitted parameters using the ``offset method''.  To apply the ``offset method'', the fit is repeated for each source of systematic uncertainty, shifting the nominal data by the uncertainty attributed to that one source.  The differences between the values of the parameters extracted from the nominal and the shifted data are then summed in quadrature and included as the total systematic uncertainty on the parameter itself. 

%%%%%%%%%%%%%%%%%%%%%%%%%%%%%%%%%%%%%%%%%%%%%%%%%%%%%%%%%%%%%%%%%%%%%%%%%%%%%%
\section{Results and discussion}
\label{sec:results}

\begin{figure}
\begin{center}
%\hspace{-2.7cm}
\includegraphics*[angle=0,scale=1.]{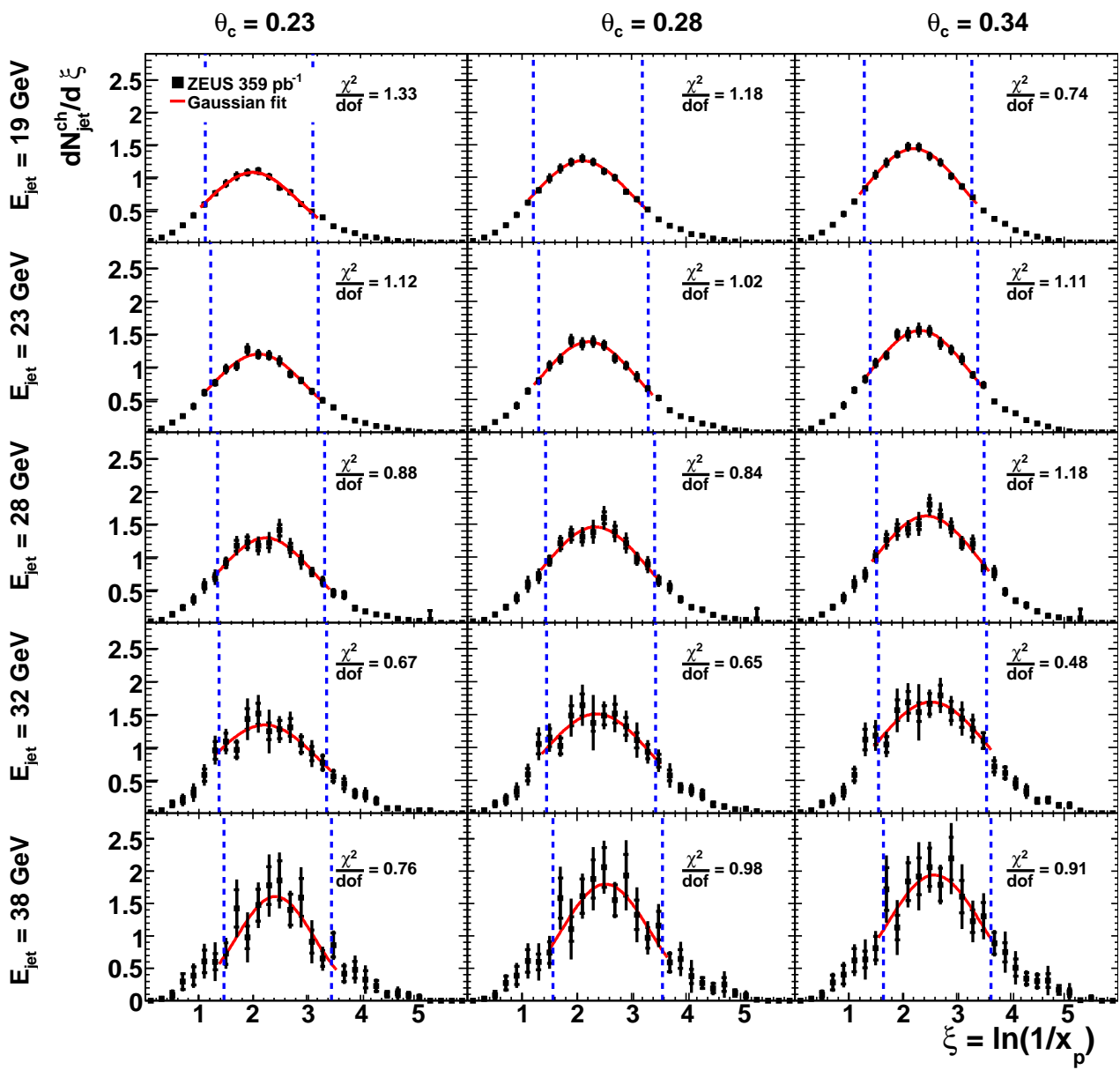}
%
%\vspace{-17.6cm}
%\hspace{0.7cm}
%\includegraphics*[angle=0,scale=0.5]{ZEUS.eps}
%\vspace{17.6cm}

\caption{
The $\xi$ distributions in the five $E_{\rm jet}$ bins using the three $\theta_c$ values. The ZEUS data are shown by the solid squares.  The inner error bars represent the statistical uncertainty.  The outer error bars represent the statistical plus systematic uncertainties added in quadrature. Gaussian functions (solid line) have been fitted to the data within the regions indicated (dashed lines). The $\chi^2/{\rm dof}$ of each fit is given on the plot.
\label{fig:fitXiGauss}}
\end{center}
\end{figure}

The $\xi$ distributions were measured in five bins of $E_{\rm jet}$ and in cones around the reconstructed jet axes with opening angles $\theta_c=\{0.23,0.28,0.34\}$.  The characteristic energy scales of the five $E_{\rm jet}$ bins, $E_{\rm jet}=\{19,23,28,32,38\}$\gev, were equated with the mean $E_{\rm jet}$ value for all events contributing to that bin.  They are shown in Fig.~\ref{fig:fitXiGauss}. Each of the distributions are observed to be similar in shape and are roughly Gaussian with more pronounced upper tails.

To assess the validity of the MLLA predictions using the measured $\xi$ distributions, two approaches were adopted.  The first, discussed in Section~\ref{sec:results:peak}, was based solely on the position of the peak of the $\xi$ distributions, $\xi_{\rm peak}$.  The second was based on the full shape of the $\xi$ distributions and is discussed in Section~\ref{sec:results:shape}.

\subsection{The \boldmath $\xi_{\rm peak}$ analysis}
\label{sec:results:peak}

The values of $\xi_{\rm peak}$ were extracted from the $\xi$ distributions using a three-parameter Gaussian fit. In accordance with previous analyses~\cite{Derrick:1995ca, Acosta:2002gg}, the distributions were fit in the range $\mu_\xi\pm 1$, where $\mu_\xi$ is the arithmetic mean of the $\xi$ distribution over the full $\xi$ range.  The explicit ranges and $\chi^2/{\rm dof}$ values of the fits are given in Fig.~\ref{fig:fitXiGauss}. The $\chi^2/{\rm dof}$ values range between $0.48$ and $1.33$ and hence indicate that the fits are reasonable. 

Uncertainty in the $\xi_{\rm peak}$ values due to the choice of fitting range was added in quadrature to the total systematic uncertainty.  It was evaluated by changing the fit range to $\mu_\xi\pm 0.9$ and $\mu_\xi\pm 1.1$, leading maximally to a $^{+0.14}_{-1.31}\%$ systematic effect. The largest and only other source contributing more than 1\% to the systematic uncertainty was the CAL energy scale, leading to a $^{+0.58}_{-2.86}\%$ effect.  The extracted values of $\xi_{\rm peak}$ are given in Table~\ref{Table_Gaussian_Peaks} and are observed to increase as the energy scale or $\theta_c$ increases.

\begin{figure}
\begin{center}
%\vspace{-1.0cm}
%\hspace{0.7cm}
%\includegraphics*[angle=0,scale=0.4]{ZEUS.eps}
%\vspace{0.0cm}
%\hspace{-0.7cm}

\includegraphics*[angle=0,scale=1.]{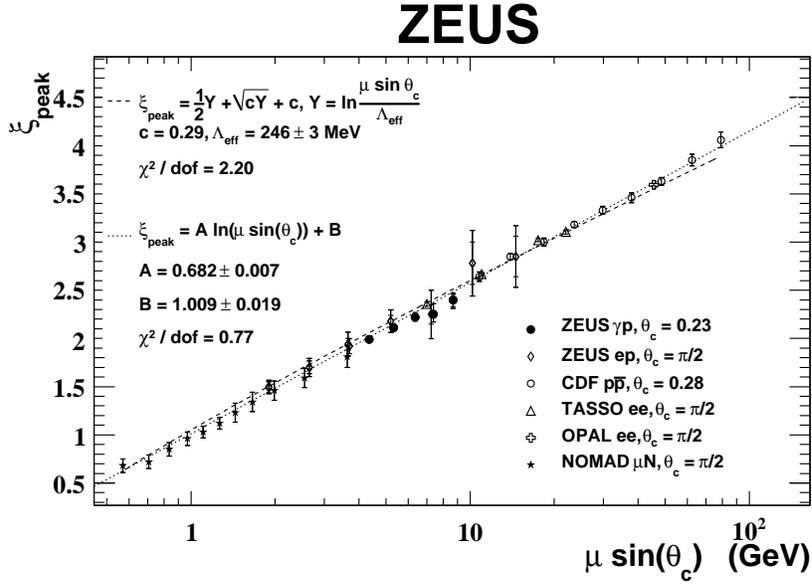}
\caption{
$\xi_{\rm peak}$ as a function of $\mu\sin(\theta_c)$, where $\mu$ denotes the characteristic energy scale for each specific process. The ZEUS $\gamma p$ data (solid circles) are shown along with $ep$ data from the ZEUS collaboration (diamonds) and results reported by the OPAL (crosses), TASSO (triangles), NOMAD (stars) and CDF (open circles) collaborations.  The inner error bars on the ZEUS points represent the statistical uncertainty.  The outer error bars represent the statistical plus systematic uncertainties added in quadrature for all data sets. The data have been fitted with a straight line.
\label{fig:peaksALL}}
\end{center}
\end{figure}

The $\xi_{\rm peak}$ values are shown in Fig.~\ref{fig:peaksALL} as a function of $\mu\sin\left(\theta_{c}\right)$, where the characteristic energy scale here is $\mu=E_{\rm jet}$. Also shown at their characteristic energy scales are data from the ZEUS $ep$ DIS~\cite{} analysis and the OPAL~\cite{Akrawy:1990ha}, TASSO~\cite{Braunschweig:1990yd}, NOMAD~\cite{Altegoer:1998py} and CDF~\cite{Acosta:2002gg} collaborations. There is an approximately linear relationship between $\xi_{\rm peak}$ and $\ln\left(E_{\rm jet}\sin\left(\theta_{c}\right)\right)$.   This relationship was tested by fitting the $\xi_{\rm peak}$ data, measured with $\theta_c=0.23$, with a straight line, parameterised as $\xi_{\rm peak} = A\left(\ln(E_{\rm jet}\sin(\theta_c))\right) + B$.  In the case where only the new ZEUS $\gamma p$ data were considered, the best fit values for the coefficients were found to be $A=0.56 \pm 0.06$(stat.)$^{+0.08}_{-0.03}$(syst.)and $B=1.16 \pm 0.09$(stat.)$^{+0.06}_{-0.14}$(syst.).  The $\chi^2/{\rm dof}$ of the fit was 0.51.  

A test of the same linear relationship was made using the global data set in Fig.~\ref{fig:peaksALL}.  The best global fit values for the coefficients were found to be $A=0.682 \pm 0.007$($\rm stat.\oplus syst.$) and $B=1.009 \pm 0.019$($\rm stat.\oplus syst.$), with a $\chi^2/{\rm dof}$ of 0.77.  Here, all systematic uncertainties were treated as uncorrelated. The globally-extracted parameters are consistent with those extracted from the ZEUS data alone. The ZEUS $\gamma p$ points are systematically below the global-fit line, however the differences are within the total experimental uncertainty. 

The MLLA in fact predicts a small square-root correction to the perfect linear dependence, as seen in Eq.~\ref{equ:2a}. Assuming $\Lambda_{\rm eff}$ is constant within the range of energies probed, Eq.~\ref{equ:2a} can be directly fit to the $\xi_{\rm peak}$ data, treating $\Lambda_{\rm eff}$ as a free parameter.  In the case where only the ZEUS $\gamma p$ data with $\theta_c=0.23$ were considered, the best fit value was found to be $\Lambda_{\rm eff}=275 \pm 4$(stat.)$^{+4}_{-8}$(syst.)\mev.  The $\chi^2/{\rm dof}$ of the fit was 0.70, indicating a good fit.  When the global data set was considered, the best fit value was found to be $\Lambda_{\rm eff}=246 \pm 3$($\rm stat.\oplus syst.$)\mev.  In the global fit, all uncertainties were treated as uncorrelated. The $\chi^2/{\rm dof}$ of the fit, with this simplistic error treatment, was 2.2, indicating some discrepancy. The globally extracted value of $\Lambda_{\rm eff}$ is not consistent with that extracted from the ZEUS data alone. 

\begin{figure}
\begin{center}
%\vspace{0.0cm}
%\hspace{0.7cm}
%\includegraphics*[angle=0,scale=0.4]{ZEUS.eps}
%\vspace{0.0cm}
%\hspace{-0.7cm}

\includegraphics*[angle=0,scale=1.]{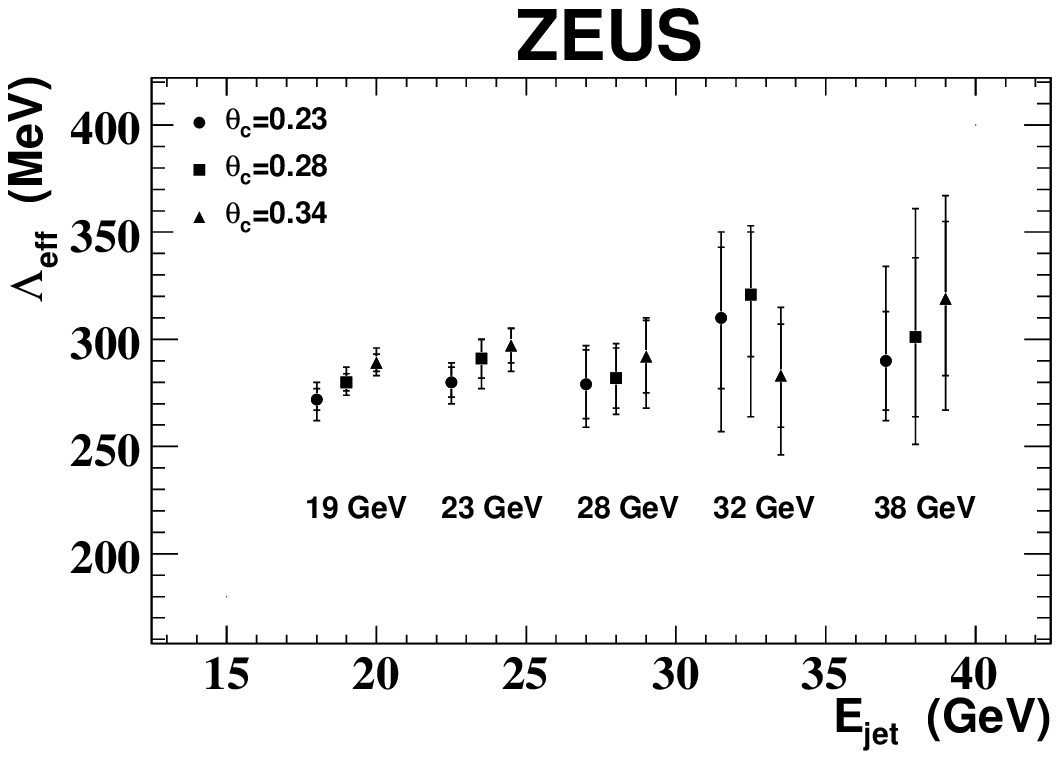}
\caption{
$\Lambda_{\rm eff}$ extracted at the five $E_{\rm jet}$ points using the three $\theta_c$ values.  The ZEUS data are shown by the solid points.  The inner error bars represent the statistical uncertainty.  The outer error bars represent the statistical plus systematic uncertainties added in quadrature. The points have been shifted horizontally for clarity.
\label{fig:LamdaZEUS}}
\end{center}
\end{figure}

The energy dependence of $\Lambda_{\rm eff}$ was studied by using Eq.~\ref{equ:2a} to map each $\xi_{\rm peak}$ value to a corresponding value of $\Lambda_{\rm eff}$.  The results, given in Table~\ref{Table_MLLA_Lambda} and shown in Fig.~\ref{fig:LamdaZEUS} as a function of $E_{\rm jet}$, show no evidence that $\Lambda_{\rm eff}$ is dependent on the energy scale.  A weak dependence was observed in the CDF data~\cite{Acosta:2002gg}, which span a wider range of energy scales.  However, the data do suggest that the value of $\Lambda_{\rm eff}$ is weakly dependent on $\theta_c$. Specifically, Fig.~\ref{fig:LamdaZEUS} shows that the values of $\Lambda_{\rm eff}$ extracted from the wider cone data tend to be systematically larger.  This behaviour was also observed by the CDF collaboration~\cite{Acosta:2002gg}.  Both the $\theta_c$ and $E_{\rm jet}$ dependence seen by CDF would contribute to the discrepancy observed when fitting Eq.~\ref{equ:2a} to the global data set.
%, which is significant for the two lowest $E_{\rm jet}$ bins when the high degree of statistical correlation between the three $\theta_c$ samples and the bin-to-bin correlation in the systematic uncertainties is taken into consideration.  

\begin{figure}
\begin{center}
%\vspace{-0.5cm}
%\hspace{0.7cm}
%\includegraphics*[angle=0,scale=0.4]{ZEUS.eps}
%\vspace{0.0cm}
%\hspace{-0.7cm}

\includegraphics*[angle=0,scale=1.]{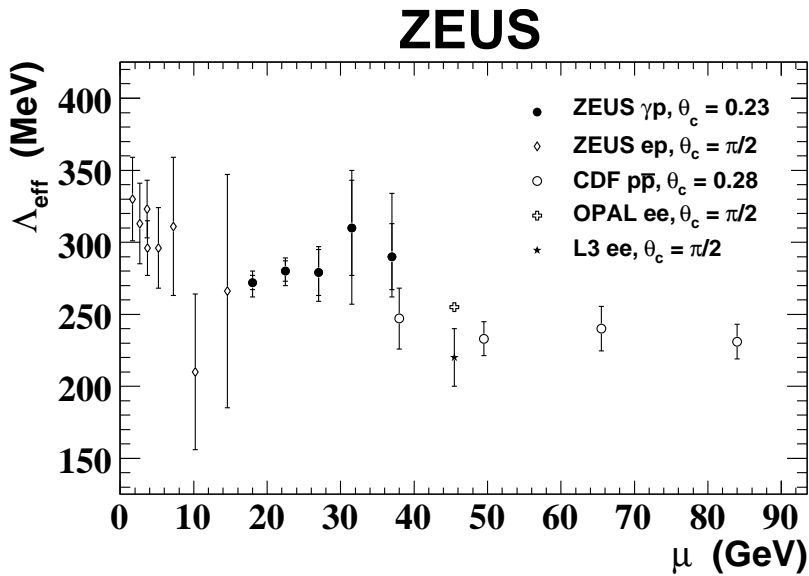}
\caption{
$\Lambda_{\rm eff}$ as a function of $\mu$, where $\mu$ denotes the characteristic energy scale for each specific process. The ZEUS $\gamma p$ data are shown by the solid circles. Also shown are $ep$ data from the ZEUS collaboration and results reported by the OPAL, L3 and CDF collaborations.  The inner error bars on the ZEUS $\gamma p$ points represent the statistical uncertainty.  The outer error bars represent the statistical plus systematic uncertainties added in quadrature for all data sets.
\label{fig:LambdaALL}}
\end{center}
\end{figure}

In Fig.~\ref{fig:LambdaALL}, the values of $\Lambda_{\rm eff}$ extracted using the $\xi_{\rm peak}$ data are shown as a function of the energy scale and compared to the previous results from ZEUS~\cite{Derrick:1995ca} using $ep$ DIS collisions, and the OPAL~\cite{Akrawy:1990ha}, L3~\cite{Adeva:1991it} and CDF~\cite{Acosta:2002gg} collaborations.  The values are all largely consistent in the energy scale region shown, supporting the prediction that $\Lambda_{\rm eff}$ is a universal parameter.

\subsection{The \boldmath $\xi$-shape analysis}
\label{sec:results:shape}

\begin{figure}
\begin{center}
%\hspace{-1.5cm}
\includegraphics*[angle=0,scale=1.]{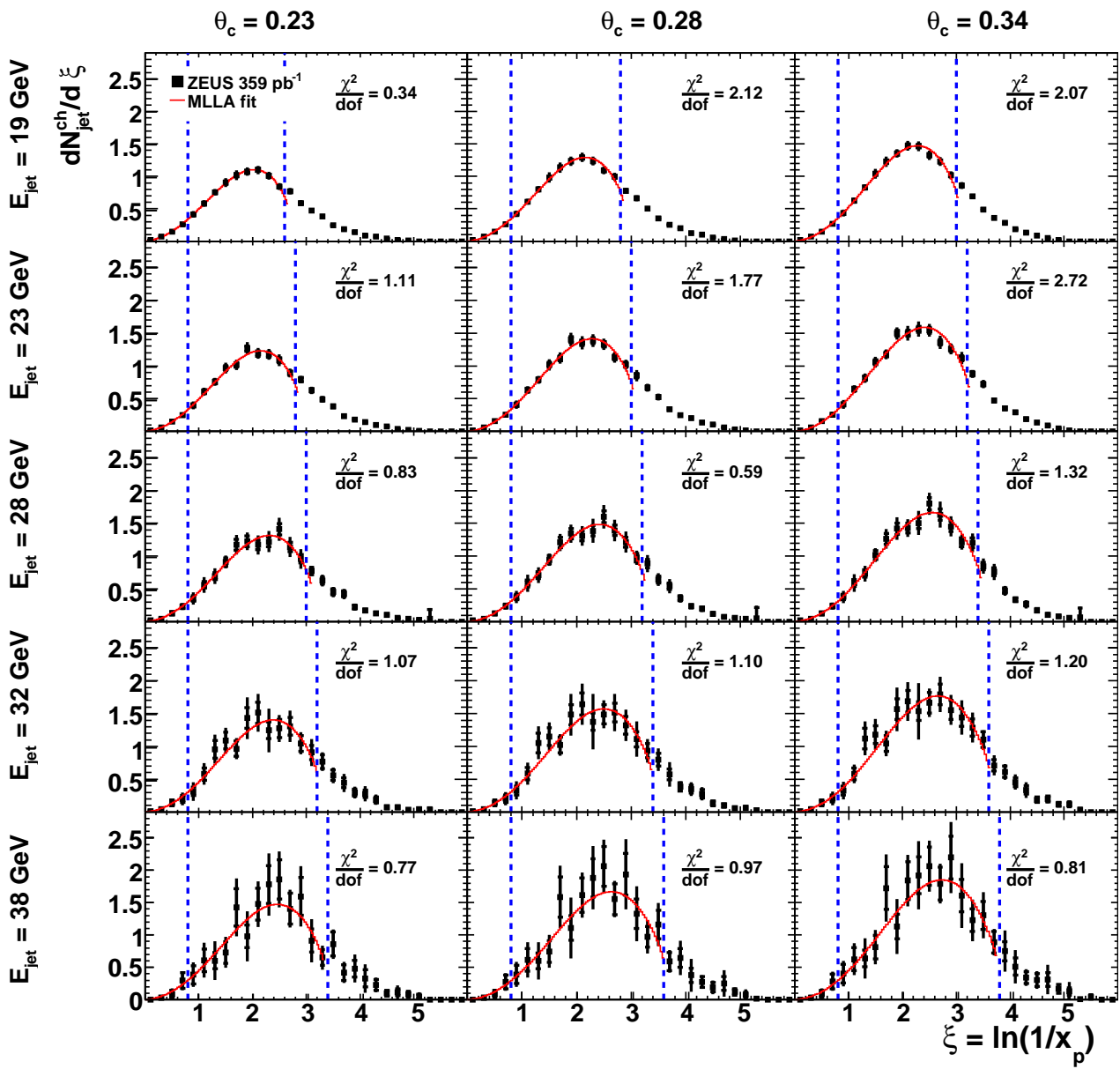}

%\vspace{-17.6cm}
%\hspace{1.5cm}
%\includegraphics*[angle=0,scale=0.5]{ZEUS.eps}
%\vspace{17.6cm}

\caption{
The $\xi$ distributions in the five $E_{\rm jet}$ bins using the three $\theta_c$ values. The ZEUS data are shown by the solid squares.  The inner error bars represent the statistical uncertainty.  The outer error bars represent the statistical plus systematic uncertainties added in quadrature. The limited momentum spectrum predicted by the MLLA (solid line) has been fitted to the data within the regions indicated (dashed lines). The $\chi^2/{\rm dof}$ of each fit is given on the plot.
\label{fig:fitXiMLLA}}
\end{center}
\end{figure}

The $\xi$ distributions were also fitted using the predicted limiting spectrum, according to Eq.~\ref{equ:5}.  The quantities $K$ and $\Lambda_{\rm eff}$ were treated as free parameters during the fit. The fitted MLLA functions are shown in Fig.~\ref{fig:fitXiMLLA}.  The fits were restricted to the ranges indicated by the vertical lines and the $\chi^2/{\rm dof}$ values of the fits are also given and lie between $0.34$ and $2.72$. Typically, in each $E_{\rm jet}$ bin, the $\chi^2/{\rm dof}$ increases as $\theta_c$ does.  The $\chi^2/{\rm dof}$ values indicate that, while the theory does describe many of the features of the data in the fitting ranges, there are differences.  Specifically, the rising edges of the $\xi$ peaks are well described. However, the upper tails of the distributions are not adequately reproduced. The same was observed in $e^+e^-$~\cite{Akrawy:1990ha,Adeva:1991it} and $ep$ DIS~\cite{Derrick:1995ca} data and to a lesser extent in high-$E_{\rm jet}$ $p\bar{p}$ data~\cite{Acosta:2002gg}.  This is likely due to the specific MLLA regularisation scheme used here and in the other aforementioned analyses. 

As discussed in Section~\ref{sec:MLLAmethod}, the MLLA regularisation scheme used here causes the partons to be cut-off at $p_T^{\rm rel,pl}=\Lambda_{\rm eff}$, whereas the hadrons in the data are not.  This leads to an intrinsic discrepancy between data and theory.  The discrepancy is present for all $\xi>0$, however the magnitude of the effect is small at low $\xi$ and increases until, for all $\xi>\ln\left(E_{\rm jet}\sin(\theta_c)/\Lambda_{\rm eff}\right)$, there are only hadrons and no partons.  

A consequence of this discrepancy is that, in order to fit the data using Eq.~\ref{equ:5}, a relatively arbitrary upper fitting bound, $\xi_+$, had to be chosen for each $\xi$ distribution. The criteria used to set $\xi_+$ were that the resulting fits were reasonably stable and that $\xi_{\rm peak}\ll \xi_+<\ln\left(E_{\rm jet}/250\mev\right)$ was satisfied, where $250$~$\mev$ roughly corresponds to the values of $\Lambda_{\rm eff}$ extracted from the $\xi_{\rm peak}$ data. The finite experimental $\xi$ binning was also a consideration. It was chosen to use $\xi_+ = w\xi_{\rm peak} + (1-w)\ln\left(E_{\rm jet}/250\mev\right)$, with $w=0.25$ for the nominal fits.  The sensitivity of $K$ and $\Lambda_{\rm eff}$ to the choice of the fitting range was treated as a systematic uncertainty and was evaluated by varying $w$ by $\pm0.1$. This source of uncertainty strongly dominates the overall uncertainty on $\Lambda_{\rm eff}$, leading to a $^{+1.8}_{-10.6}\%$ effect, although $K$ was found to be largely insensitive to it.  The same lower fitting bound, $\xi_-=\ln(2)$, was used is all cases and both $K$ and $\Lambda_{\rm eff}$ were observed to be insensitive to a variation of $\xi_-$ by $\pm15$\%.  

The values of $\Lambda_{\rm eff}$ extracted from the MLLA fits are given in Table~\ref{Table_MLLA_Lambda}. The results are in reasonable agreement with those extracted from the $\xi_{\rm peak}$ data, although the values extracted using the MLLA fit have larger uncertainties. The value of $\Lambda_{\rm eff}$ from the MLLA method with $\theta_c=0.23$ and averaged over $E_{\rm jet}$, weighting each data point based only on its statistical precision, is $\Lambda_{\rm eff}= 304 \pm 6$(stat.)$^{+8}_{-32}$(syst.)$\mev$.

\begin{figure}
\begin{center}
%\vspace{-0.5cm}
%\hspace{0.7cm}
%\includegraphics*[angle=0,scale=0.4]{ZEUS.eps}
%\vspace{0.0cm}
%\hspace{-0.7cm}

\includegraphics*[angle=0,scale=1.]{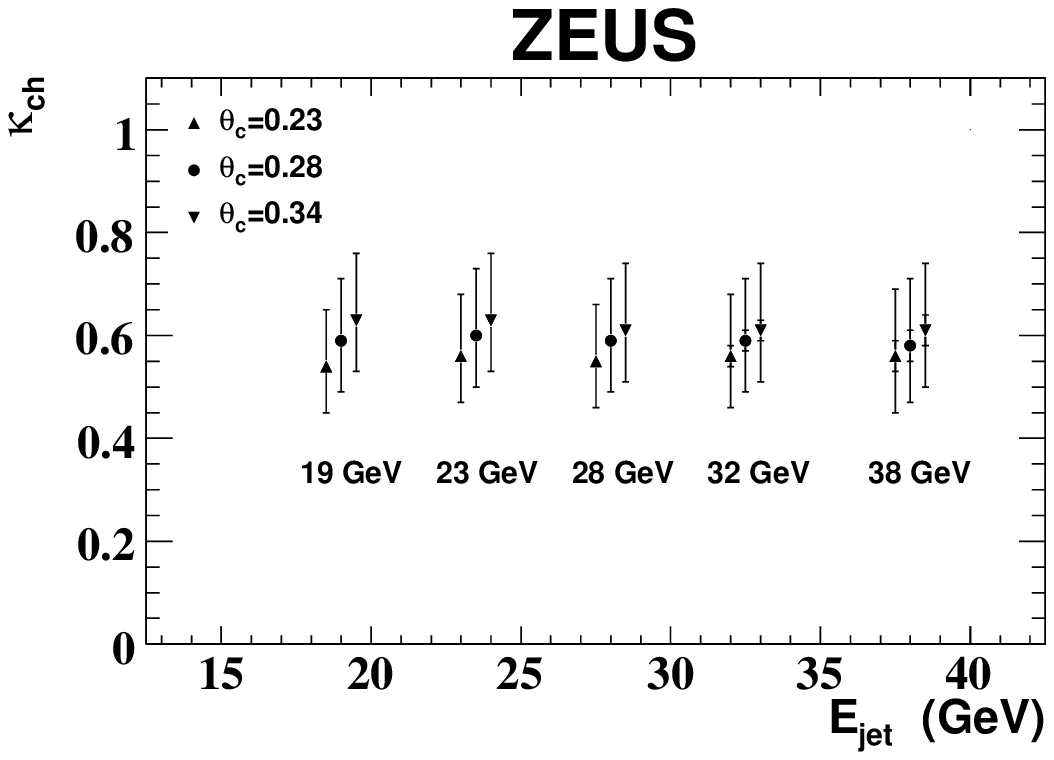}
\caption{
$\kappa^{\rm ch}$ extracted at the five $E_{\rm jet}$ points using the three $\theta_c$ values.  The ZEUS data are shown by the solid points.  The inner error bars represent the statistical uncertainty.  The outer error bars represent the statistical, systematic and theoretical uncertainties added in quadrature. The points have been shifted horizontally for clarity.
\label{fig:kappaZEUS}}
\end{center}
\end{figure}

Values of $\kappa_{\rm ch}$ were extracted from the fitted $K$ values using Eq.~\ref{equ:5} and the values of $\epsilon_{\rm g}$ predicted for each $E_{\rm jet}$ bin by the {\sc Pythia} model. The $\epsilon_{\rm g}$ values were roughly constant in $E_{\rm jet}$, at $\epsilon_{\rm g}\approx0.2$.  The $\kappa_{\rm ch}$ values are given in Table~\ref{Table_MLLA_Kappa} and are shown in Fig.~\ref{fig:kappaZEUS}.  The total uncertainty is dominated by the theoretical uncertainty associated with the next-to-MLLA correction factors.  The $\kappa_{\rm ch}$ data suggest a weak dependence on $\theta_c$.  Specifically, as $\theta_c$ increases, so too does the central value of $\kappa_{\rm ch}$. This is significant when the high degree of statistical correlation between the three $\theta_c$ samples and the bin-to-bin correlation in the systematic and theoretical uncertainties are taken into consideration. The same is true for the $\kappa_{\rm ch}$ values reported by the CDF collaboration~\cite{Acosta:2002gg}, which were obtained using a different extraction method.  The ZEUS data in Fig.~\ref{fig:kappaZEUS} do not provide any evidence that $\kappa_{\rm ch}$ is dependent on $E_{\rm jet}$ in the range probed. 

The value of $\kappa_{\rm ch}$, measured with $\theta_c=0.23$ and averaged over $E_{\rm jet}$, weighting the data points based on their statistical precision, was $\kappa_{\rm ch}=0.55 \pm 0.01$(stat.)$^{+0.03}_{-0.02}$(syst.)$^{+0.11}_{-0.09}$(theo.).  The $\kappa_{\rm ch}$ value extracted here is in good agreement with that reported by the CDF collaboration, $\kappa_{\rm ch}=0.56 \pm 0.05$(stat.)$\pm 0.09$(syst.).  To compare to the values extracted using $e^+e^-$ and $ep$ DIS data and assuming no contamination from gluon jets, the values have to be scaled by $rC_F/F_{\rm nMLLA}C_A\approx 0.55$. This leads to values of $\kappa_{\rm ch}\approx 0.7$.  These other results were found with $\theta_c$ effectively set to $\pi/2$ however.

%%%%%%%%%%%%%%%%%%%%%%%%%%%%%%%%%%%%%%%%%%%%%%%%%%%%%%%%%%%%%%%%%%%%%%%%%%%%%%

\section{Summary}

The multiplicity distributions of charged particles within cones centred on jets have been measured as a function of $\xi=\ln\left(1/x_p\right)$, where $x_p$ is the fraction of the jet's momentum carried by the charged particle. These $\xi$ distributions have been measured in five bins of $E_{\rm jet}$ and with three different cone opening angles, $\theta_c$, for $\gamma p$ events containing two and only two jets, using 359 pb$^{-1}$ of $ep$ data.

The peak positions of the $\xi$ distributions, $\xi_{\rm peak}$, were extracted and observed to increase roughly linearly with $\ln\left(E_{\rm jet}\sin\left(\theta_{c}\right)\right)$.  A single value of intrinsic MLLA scale, $\Lambda_{\rm eff}$, was extracted by fitting the $\xi_{\rm peak}$ data according to the predicted relationship between $\xi_{\rm peak}$ and $\ln\left(E_{\rm jet}\sin\left(\theta_{c}\right)/\Lambda_{\rm eff}\right)$.  The best fit value was found to be $\Lambda_{\rm eff}=275 \pm 4$(stat.)$^{+4}_{-8}$(syst.)\mev. 

The $E_{\rm jet}$ and $\theta_c$ dependences of $\Lambda_{\rm eff}$ were studied by calculating a value of $\Lambda_{\rm eff}$ from each $\xi_{\rm peak}$ data point. The value of $\Lambda_{\rm eff}$ weakly depends on $\theta_c$ but no $E_{\rm jet}$ dependence was observed. The $\Lambda_{\rm eff}$ data are consistent with previously published data sets using different initial states, supporting the prediction that $\Lambda_{\rm eff}$ is universal.

The $\xi$ distributions were also fitted using the limited momentum spectra predicted by the MLLA and assuming LPHD, in the regions where they are applicable.  The theory largely described the data in these regions.  The fitted MLLA functions were used to extract the value of $\Lambda_{\rm eff}$ as a function of $E_{\rm jet}$ and $\theta_c$.  The value extracted using this method with $\theta_c=0.23$ and averaged over $E_{\rm jet}$, was $\Lambda_{\rm eff}= 304 \pm 6$(stat.)$^{+8}_{-32}$(syst.)$\mev$.

The value of the LPHD parameter $\kappa_{\rm ch}$ was extracted as a function of $E_{\rm jet}$ and $\theta_c$ from the fitted limited momentum spectra. Corrections based on next-to-MLLA theory were included. The value extracted with $\theta_c=0.23$ and averaged over $E_{\rm jet}$, was $\kappa_{\rm ch}=0.55 \pm 0.01$(stat.)$^{+0.03}_{-0.02}$(syst.)$^{+0.11}_{-0.09}$(theo.).  The value of $\kappa_{\rm ch}$ has a weak dependence on $\theta_c$ and is consistent with the results published by the CDF collaboration. The data support the assumption that $\kappa_{\rm ch}$ is universal.

\section{Acknowledgments}

We would like to sincerely thank Wolfgang Ochs for many highly illuminating conversations.  We appreciate the contributions to the construction and maintenance of the ZEUS detector of many people who are not listed as authors.  The HERA machine group and the DESY computing staff are especially acknowledged for their success in providing excellent operation of the collider and the data analysis environment.  We thank the DESY directorate for their strong support and encouragement.

{
\def\bibname{\Large\bf References}
\def\refname{\Large\bf References}
\pagestyle{plain}
\ifzeusbst
  \bibliographystyle{./BiBTeX/bst/l4z_default}
\fi
\ifzdrftbst
  \bibliographystyle{./BiBTeX/bst/l4z_draft}
\fi
\ifzbstepj
  \bibliographystyle{./BiBTeX/bst/l4z_epj}
\fi
\ifzbstnp
  \bibliographystyle{./BiBTeX/bst/l4z_np}
\fi
\ifzbstpl
  \bibliographystyle{./BiBTeX/bst/l4z_pl}
\fi
{\raggedright
\bibliography{./BiBTeX/user/syn.bib,%
              ./BiBTeX/bib/l4z_articles.bib,%
              ./BiBTeX/bib/l4z_books.bib,%
              ./BiBTeX/bib/l4z_conferences.bib,%
              ./BiBTeX/bib/l4z_h1.bib,%
              ./BiBTeX/bib/l4z_misc.bib,%
              ./BiBTeX/bib/l4z_old.bib,%
              ./BiBTeX/bib/l4z_preprints.bib,%
              ./BiBTeX/bib/l4z_replaced.bib,%
              ./BiBTeX/bib/l4z_temporary.bib,%
              ./BiBTeX/bib/l4z_zeus.bib}}
}
\vfill\eject

%%%%%%%%%%%%%%%%%%%%%%%%%%%%%%%%%%%%%%%%%%%%%%%%%%%%%%%%%%%%%%%%%%%%%%%%%%%

\begin{table}
\centering
\begin{tabular}{|c|c||c|c|c|}
\hline
$E_{\rm jet}$ ($\gev$) & $\theta_c$ & $\xi_{\rm peak}$ & stat. & syst. \\
\hline
\hline
\multirow{3}{*}{19} 
& 0.23 & 1.99 & $\pm0.01$ & $^{+0.02}_{-0.02}$ \\   
& 0.28 & 2.10 & $\pm0.01$ & $^{+0.01}_{-0.01}$ \\   
& 0.34 & 2.20 & $\pm0.01$ & $^{+0.01}_{-0.01}$ \\   
\hline
\multirow{3}{*}{23} 
& 0.23 & 2.11 & $\pm0.02$ & $^{+0.02}_{-0.01}$ \\   
& 0.28 & 2.21 & $\pm0.02$ & $^{+0.02}_{-0.01}$ \\   
& 0.34 & 2.32 & $\pm0.02$ & $^{+0.02}_{-0.01}$ \\   
\hline
\multirow{3}{*}{28} 
& 0.23 & 2.22 & $\pm0.04$ & $^{+0.03}_{-0.02}$ \\   
& 0.28 & 2.34 & $\pm0.03$ & $^{+0.02}_{-0.02}$ \\   
& 0.34 & 2.44 & $\pm0.04$ & $^{+0.04}_{-0.01}$ \\   
\hline
\multirow{3}{*}{32} 
& 0.23 & 2.25 & $\pm0.07$ & $^{+0.09}_{-0.05}$ \\   
& 0.28 & 2.36 & $\pm0.06$ & $^{+0.10}_{-0.03}$ \\   
& 0.34 & 2.56 & $\pm0.06$ & $^{+0.07}_{-0.05}$ \\   
\hline
\multirow{3}{*}{38} 
& 0.23 & 2.40 & $\pm0.05$ & $^{+0.04}_{-0.08}$ \\   
& 0.28 & 2.50 & $\pm0.08$ & $^{+0.07}_{-0.18}$ \\   
& 0.34 & 2.59 & $\pm0.07$ & $^{+0.08}_{-0.15}$ \\   
\hline
\end{tabular}
\caption{$\xi_{\rm peak}$ values in the five $E_{\rm jet}$ bins using the three $\theta_c$ values.  The statistical and systematic uncertainties are also given.}
\label{Table_Gaussian_Peaks} 
\end{table}

%%%%%%%%%%%%%%%%%%%%%%%%%%%%%%%%%%%%%%%%%%%%%%%%%%%%%%%%%%%%%%%%%%%%%%%%%%%

\begin{table}
\centering
\begin{tabular}{|c|c||c|c|c||c|c|c|}
\hline
\multicolumn{2}{|c||}{} & \multicolumn{3}{c||}{$\xi_{\rm peak}$ analysis} & \multicolumn{3}{c|}{$\xi$ shape analysis} \\
\hline
$E_{\rm jet}$ ($\gev$) & $\theta_c$ 
& $\Lambda_{\rm eff}$ ($\mev$) & stat. & syst.
& $\Lambda_{\rm eff}$ ($\mev$) & stat. & syst. \\
\hline
\hline
\multirow{3}{*}{19} 
& 0.23 & 272 & $\pm5$ & $^{+6}_{-8}$ & 304 & $\pm4$ & $^{+7}_{-32}$  \\
& 0.28 & 280 & $\pm4$ & $^{+5}_{-5}$ & 298 & $\pm4$ & $^{+21}_{-25}$ \\
& 0.34 & 289 & $\pm4$ & $^{+6}_{-5}$ & 303 & $\pm3$ & $^{+15}_{-30}$ \\
\hline
\multirow{3}{*}{23} 
& 0.23 & 280 & $\pm7$ & $^{+6}_{-7}$  & 307 & $\pm6$ & $^{+10}_{-32}$ \\
& 0.28 & 291 & $\pm9$ & $^{+3}_{-11}$ & 305 & $\pm6$ & $^{+23}_{-32}$ \\
& 0.34 & 297 & $\pm8$ & $^{+3}_{-9}$  & 301 & $\pm5$ & $^{+26}_{-29}$ \\
\hline
\multirow{3}{*}{28} 
& 0.23 & 279 & $\pm16$ & $^{+8}_{-11}$ & 285 & $\pm12$ & $^{+8}_{-19}$  \\
& 0.28 & 282 & $\pm14$ & $^{+8}_{-9}$  & 294 & $\pm10$ & $^{+7}_{-29}$  \\
& 0.34 & 292 & $\pm17$ & $^{+5}_{-17}$ & 287 & $\pm9$  & $^{+29}_{-23}$ \\
\hline
\multirow{3}{*}{32} 
& 0.23 & 310 & $\pm33$ & $^{+22}_{-41}$ & 298 & $\pm15$ & $^{+25}_{-40}$ \\
& 0.28 & 321 & $\pm29$ & $^{+14}_{-49}$ & 302 & $\pm13$ & $^{+26}_{-41}$ \\
& 0.34 & 283 & $\pm24$ & $^{+21}_{-28}$ & 286 & $\pm14$ & $^{+28}_{-27}$ \\
\hline
\multirow{3}{*}{38} 
& 0.23 & 290 & $\pm23$ & $^{+38}_{-16}$ & 311 & $\pm15$ & $^{+13}_{-52}$ \\
& 0.28 & 301 & $\pm37$ & $^{+48}_{-33}$ & 287 & $\pm21$ & $^{+42}_{-32}$ \\
& 0.34 & 319 & $\pm36$ & $^{+31}_{-38}$ & 297 & $\pm17$ & $^{+21}_{-42}$ \\
\hline
\end{tabular}
\caption{$\Lambda_{\rm eff}$ extracted at the five $E_{\rm jet}$ points using the three $\theta_c$ values obtained from both the $\xi_{\rm peak}$ and $\xi$ shape analyses. The statistical and systematic uncertainties are also given.}
\label{Table_MLLA_Lambda}
\end{table}

%%%%%%%%%%%%%%%%%%%%%%%%%%%%%%%%%%%%%%%%%%%%%%%%%%%%%%%%%%%%%%%%%%%%%%%%%%%

\begin{table}
\centering
\begin{tabular}{|c|c||c|c|c|c|}
\hline
$E_{\rm jet}$ ($\gev$) & $\theta_c$ & $\kappa^{\rm ch}$ & stat. & syst. & theo. \\
\hline
\hline
\multirow{3}{*}{19} 
& 0.23 & 0.54 & $\pm0.01$ & $^{+0.03}_{-0.02}$ & $^{+0.11}_{-0.09}$ \\   
& 0.28 & 0.59 & $\pm0.01$ & $^{+0.03}_{-0.01}$ & $^{+0.12}_{-0.10}$ \\   
& 0.34 & 0.63 & $\pm0.01$ & $^{+0.03}_{-0.02}$ & $^{+0.12}_{-0.10}$ \\   
\hline
\multirow{3}{*}{23} 
& 0.23 & 0.56 & $\pm0.01$ & $^{+0.03}_{-0.02}$ & $^{+0.11}_{-0.09}$ \\   
& 0.28 & 0.60 & $\pm0.01$ & $^{+0.04}_{-0.02}$ & $^{+0.12}_{-0.10}$ \\   
& 0.34 & 0.63 & $\pm0.01$ & $^{+0.04}_{-0.02}$ & $^{+0.13}_{-0.10}$ \\   
\hline
\multirow{3}{*}{28} 
& 0.23 & 0.55 & $\pm0.01$ & $^{+0.04}_{-0.01}$ & $^{+0.11}_{-0.09}$ \\   
& 0.28 & 0.59 & $\pm0.01$ & $^{+0.04}_{-0.04}$ & $^{+0.11}_{-0.09}$ \\   
& 0.34 & 0.61 & $\pm0.01$ & $^{+0.04}_{-0.02}$ & $^{+0.12}_{-0.10}$ \\   
\hline
\multirow{3}{*}{32} 
& 0.23 & 0.56 & $\pm0.02$ & $^{+0.04}_{-0.04}$ & $^{+0.11}_{-0.09}$ \\   
& 0.28 & 0.59 & $\pm0.02$ & $^{+0.04}_{-0.04}$ & $^{+0.11}_{-0.09}$ \\   
& 0.34 & 0.61 & $\pm0.02$ & $^{+0.04}_{-0.03}$ & $^{+0.12}_{-0.10}$ \\   
\hline
\multirow{3}{*}{38} 
& 0.23 & 0.56 & $\pm0.03$ & $^{+0.05}_{-0.06}$ & $^{+0.11}_{-0.09}$ \\   
& 0.28 & 0.58 & $\pm0.03$ & $^{+0.04}_{-0.04}$ & $^{+0.11}_{-0.09}$ \\   
& 0.34 & 0.61 & $\pm0.03$ & $^{+0.03}_{-0.05}$ & $^{+0.12}_{-0.10}$ \\   
\hline
\end{tabular}
\caption{$\kappa^{\rm ch}$ values extracted at the five $E_{\rm jet}$ points using the three $\theta_c$ values.  The statistical, systematic and theoretical uncertainties are also given.}
\label{Table_MLLA_Kappa}
\end{table}

%%%%%%%%%%%%%%%%%%%%%%%%%%%%%%%%%%%%%%%%%%%%%%%%%%%%%%%%%%%%%%%%%%%%%%%%%

%%%%%%%%%%%%%%%%%%%%%%%%%%%%%%%%%%%%%%%%%%%%%%%%%%%%%%%%%%%%%%%%%%%%%%%%%

%%%%%%%%%%%%%%%%%%%%%%%%%%%%%%%%%%%%%%%%%%%%%%%%%%%%%%%%%%%%%%%%%%%%%%%%%

%%%%%%%%%%%%%%%%%%%%%%%%%%%%%%%%%%%%%%%%%%%%%%%%%%%%%%%%%%%%%%%%%%%%%%%%%

%%%%%%%%%%%%%%%%%%%%%%%%%%%%%%%%%%%%%%%%%%%%%%%%%%%%%%%%%%%%%%%%%%%%%%%%%

%%%%%%%%%%%%%%%%%%%%%%%%%%%%%%%%%%%%%%%%%%%%%%%%%%%%%%%%%%%%%%%%%%%%%%%%%

%%%%%%%%%%%%%%%%%%%%%%%%%%%%%%%%%%%%%%%%%%%%%%%%%%%%%%%%%%%%%%%%%%%%%%%%%

\end{document}